\documentstyle[12pt]{article}
\newcommand{\be}{\begin{equation}}
\newcommand{\ee}{\end{equation}}
\begin{document}
\begin{titlepage}
\title{ 
{\bf K-system generator of pseudorandom numbers on Galois field
\footnote{Reprint DEMO-HEP 97/03 Feb.97}} 
}
{\bf 
\author{ 
G.G.Athanasiu\\
Physics Department,University of Crete\\
GR-71409 Iraklion, Crete, Greece\\
  \vspace{1cm}\\
E.G.Floratos\\
National Research Center "Demokritos",\\
GR-15310 Ag. Paraskevi, Athens, Greece;\\
Physics Department,University of Crete,\\
GR-71409 Iraklion,Greece\\
\vspace{1cm}\\
G.K. Savvidy\\ 
National Research Center "Demokritos",\\
GR-15310 Ag. Paraskevi, Athens, Greece
}
} 
\date{}
\maketitle 
\begin{abstract} 
\noindent

We analyze the structure of the periodic trajectories of the K-system
generator of pseudorandom numbers on a rational sublattice which coincides with 
the Galois field $GF[p]$. The period of the trajectories 
increases as a function of the lattice size $p$ and the dimension of the 
K-matrix $d$. We emphasize the connection of this approach with the one
which is based on primitive matrices over Galois fields.
\end{abstract} 
\thispagestyle{empty} 
\end{titlepage} 
\pagestyle{empty}

\section{Introduction}

Nowadays the Monte-Carlo method has a wide range of applications and the 
quality of pseudorandom numbers being used plays an important role.
Different principles and algorithms have been suggested in the literature
to generate pseoudorandom numbers and to check their properties \cite{knuth}.
The development of the ergodic theory \cite{kolmogorov,anosov,rohlin,sinai}
and the progress in understanding 
of nonlinear phenomena together with the increasing power of modern 
computers open a new  era  for applications \cite{creutz,amb,amb1}.

In the articles \cite{savvidy1} 
the authors suggested to use many-dimensional  Kolmogorov  K-systems
to generate pseudorandom numbers of high quality.
K-systems are the most stochastic 
dynamical systems, with nonzero Kolmogorov entropy and their 
trajectories are exponentially unstable and 
uniformly fill the phase space
\cite{kolmogorov,anosov,fomin,savvidy2,savvidy3,savvidy1}. 
It was suggested to use the coordinates 
of these trajectories as a sequence of pseudorandom numbers
\cite{savvidy1}. From this point of view  
the most successful inversive congruential 
generator \cite{knuth} used so far can be considered as a 
one-dimensional K-system and it was pointed out 
that this fact explains its exceptional properties \cite{savvidy1}. 

For the application of this idea
it is important to have  such K-systems
for which the phase space is limited by a unit $d$-dimensional torus,
because 
in that case the coordinates of the trajectories can be used directly
without any additional transformations. Two types of K-systems have been
suggested for these purposes: toral automorphisms \cite{savvidy1,akopov1}
and many-dimensional Sinai billiard which is defined inside a
unit $d$-dimensional torus \cite{abramyan}.

In the case of toral automorphisms a unit $d$-dimensional
torus $\Pi^{d}$ plays the role of a phase space and the K-system 
is represented by a $d$-dimensional matrix - {\it K-matrix } -
which acts on the vectors
from $\Pi^{d}$ generating trajectories uniformly distributed over the 
torus $\Pi^{d}$. The coordinates of these trajectories are used
for Monte-Caro simulations 
\cite{savvidy1,akopov1}. The properties of this new class of matrix
generators were investigated by different criterion 
including Kolmogorov discrepancy $D_{N}$ . In all 
cases it shows good statistical properties \cite{akopov1}.

The aim of this article is to estimate the period of the 
trajectories which are used to produce pseudorandom numbers 
generated by a K-system. It is clear, that  
only periodic trajectories of K-systems can be simulated on 
a computer, because trajectories on a computer are always on 
a finite rational sublattice $Z^{d}_{p}$ of the phase space $\Pi^{d}$. 
Thus we have to consider the system 
on rational sublattice $Z^{d}_{p}$ of a unit $d$-dimensional torus and 
particularly on sublattices with $prime$ basis $p$ 
\cite{barry,vivaldy1,vivaldy2,dyson,bartuccelli,pregrad,esposti,athanas}. 
These sublattices are equivalent to Galois fields 
$GF[p]$ and all four elementary arithmetical operations can be 
carried out unrestrictevely \cite{apostol,lidl,bastida}.

Analyzing trajectories of a K-system on a Galois
sublattice $GF[p]$ one can see 
that in order to have trajectories with large period
K-matrix should have an
eigenvalues in high extensions $GF[\sqrt[d]{p}]$ of
the field (notation used in mathematical literature is $GF[p^d]$). 
This property makes them very close to so called
{\it primitive matrices} which have been considered by Niedereiter
\cite{niederreiter,grothe} to generate pseoudorandom numbers.
We refer to the book of Niederreiter
\cite{nied} and to the survey article \cite{niede} for recent 
references. The main idea of his approach is to use a 
primitive matrices on a given Galois field $GF[\sqrt[d]{p}]$
to generate pseudorandom numbers of very large periods. 
This approach guarantees the large period of the series. In addition 
the fascinating result of Niedereiter \cite{niede}
allows to estimate the uniformity of maximally long trajectories 
in terms of Kolmogorov discrepancy $D_{N}$.

Thus these two approaches are very close to each 
other on Galois sublattices and the main question is: {\it whether one can 
have the matrices with 
both properties at the same time?} The  
determinant of a K-matrix should be equal to one
while the determinant 
of a primitive matrix is different from one, thus
{\it these properties are incompatible}. The main point, which we
would like to stress here, is that nevertheless one 
can construct  
K-matrices which have a {\it primitive matrix as submatrices}. 
In that case the trajectories are still very long as in
the case of primitive matrices, but at the expense of appearance 
of trajectories with short period.
Excluding them from initial data we guarantee that the trajectories are
maximally long and at the same time belong to a K-system.
We suggest specific matrices with these properties
which can be used for practical simulations.

\section{Trajectories of K-system on a rational sublattice}

Let us pass to the details of the algorithm. 
The  matrix  generator is defined as \cite{savvidy1,akopov1},

$$X^{(n+1)} =A \cdot  X^{(n)},~~~~~~~~~~~~ (mod~1),\eqno(1)$$
where $A$ is $d \times d$ dimensional matrix with integer 
matrix elements $a_{i,j}$ and determinant equal to one

$$Det~A = 1 ,\eqno(2)$$
and $X^{(0)}=(X^{(0)}_{1},...,X^{(0)}_{d})$  is an  initial real vector. 
The last condition provides phase-space volume conservation.
The automorphism (1) forms the $K$-system of Anosov if and only if 
all eigenvalues of the matrix $A$ are in modulus different 
from unity \cite{anosov,rohlin,sinai}

$$\vert \lambda_{i} \vert \neq 1,~~~~~~~i=1,...,d \eqno(3)$$ 
The $\it trajectory$ of the $K$-system (1) 

$$X_{0},X_{1},X_{2}....$$ 
represents the desired sequence of the pseudorandom numbers  
\cite{savvidy1}. 

This approach allows a 
large freedom in choosing the matrices $A$ for the K-system 
generators and the initial vectors \cite{savvidy1}.
Specific choices suggested in \cite{savvidy1,akopov1,nersesian} are

$$ A_{d} = \left( \begin{array}{c}

        2,3,4,.......,d~~,1 \\
        1,2,3,.....,d-1,1 \\
        1,1,2,.....,d-2,1 \\
        ................. \\
        ................. \\
        1,1,1,...,2,3,4,1 \\
        1,1,1,...,1,2,2,1 \\
        1,1,1,...,1,1,2,1 \\
        1,1,1,...,1,1,1,1 
\end{array} \right) , A_{d}= \left( \begin{array}{c}
         0,~~1~,~~0~,.....,~~0 \\
         0,~~0~,~~1~,.....,~~0   \\
         .............   \\
         ............. \\
         0,~~0~~,~~0~~,.....,~~1 \\
         (-1)^{d+1},a_{1},a_{2},..,a_{d-1}

\end{array} \right) . \eqno(4)$$
The first  matrix has the advantage to be well defined in any 
dimension $d$ and it has a very large Kolmogorov entropy \cite{savvidy1}
which is given by the Anosov-Sinai formula

$$
h(A_{d}) = \sum_{\vert\lambda_{k}\vert > 1} ln \lambda_{k}.
$$
The entropy $h$ defines the number $\pi (\tau)$ of the 
periodic trajectories with 
period less or equal to $\tau$ \cite{sinai5,parry,esposti}

$$ \pi(\tau) \rightarrow  \frac{e^{h \tau}}{h \tau}$$
when $\tau \rightarrow \infty$, 
thus the number of "available" trajectories increases with entropy.
The second one has a very simple expression for its characteristic 
polynomial 

$$\lambda^{d}-a_{d-1}~\lambda^{d-1}-...-a_{1}~\lambda + (-1)^{d} = $$
and for its eigenvalues $\lambda_{1},...,\lambda_{d}$
we have $\lambda_{1}\cdot \cdot \cdot \lambda_{d} =1$,...,
$\lambda_{1} +...+ \lambda_{d}= a_{d-1}$.
These formulas allow to choose eigenvalues and then to 
construct K-matrices.
This correspondence between matrices and polynomials
has a wide range of applications in algebra and number theory 
\cite{lidl}. In the given case $Det A_{d}=1$ to fulfil
K-condition (2) \cite{nersesian}. 

Let us consider trajectories of the system (1) 
with an initial vector 
$X^{(0)}$ which has rational coordinates 
\cite{barry,vivaldy1,vivaldy2,dyson,bartuccelli,esposti,athanas}

$$ X^{(0)} = (~~\frac{q_{1}}{p_{1}},~~\frac{q_{2}}{p_{2}},...,
\frac{q_{d}}{p_{d}}~~).\eqno(5)$$
It is easy to see, that all these trajectories 
are periodic orbits of the 
Anosov map (1), because matrix elements~~ $a_{i,j}$~~ are integer.
Indeed, if we consider the sublattice of unit torus $\Pi^{d}$ 
with rational coordinates of the form~~ $q/p$~~ where $~p~$ is 
the least  common multiple of p's

$$X=(~~\frac{q_{1}}{p},~~\frac{q_{2}}{p},...,
\frac{q_{d}}{p}~~), ~~~~~~0\leq q_{i} \leq p-1 $$
then the multiplication,summation and $(mod)$ operations (1) will leave the 
trajectory on the same sublattice. The total number of vertices on this 
sublattice $Z_{p}^{d}$ is 

$$(total~number~of~verteces)= p^{d},$$
therefore the period $\tau_{p}$ of the trajectories on 
$Z^{d}_{p}=Z_{p}\otimes ...\otimes Z_{p}$, where 
$Z_{p}=\{0,1,...,p-1 \}$ 
is always less than $p^{d}$

$$\tau_{p} \leq p^{d}.$$
Thus the periodic trajectories of this system (1) with the initial vector 
(5) coincide with a subset of the 
points of rational sublattice $Z^{d}_{p}$ and 
our goal is to find conditions under which the period of the 
{\it K-system} will be as large as possible.

Let us show that on every given sublattice $Z^{d}_{p}$ Anosov map (1) 
reduces to 
($\it mod$ $p$) arithmetic. Indeed on sublattice $Z^{d}_{p}$ the Anosov
map $A$ (1) can be written as

$$\frac{ q^{(n+1)}_{i}}{p} =
\sum_{j}a_{i,j}~\frac{q^{(n)}_{i}}{p},~~~~~~~~~~(mod~~1) $$ 
and is equivalent to ($mod~~p$) arithmetic on the lattice with 
integer coordinates $q_{i}$ which are in the interval $[0,p-1]$ 

$$q^{(n+1)}_{i} =\sum_{j}a_{i,j}~q^{(n)}_{i},~~~~~~~~~(mod~~p).$$
Thus the images of the 
periodic trajectories on a unit torus $\Pi^{d}$ appear as trajectories
on the integer sublattice $Z^{d}_{p}$ and all operations can be 
understood ($mod~~p$). The most important thing is that now all 
operations become commutative.

To estimate the period of the trajectories on a rational 
sublattice it is essential to consider those sublattices 
for which $p$ is the prime number, we mean that $p_{1}=...=p_{d} =
p$ \cite{barry,vivaldy1,vivaldy2,dyson,bartuccelli,esposti,athanas}. 
In that case the integer 
sublattice gains an additional structure and becomes the Galois 
field $GF[p]$ and all operations reduce to arithmetic ones on 
Galois field.
The benefit to work on Galois field is that four arithmetic 
operations are well defined on that sublattice \cite{apostol}.

In this way we can consider every coordinate $q_{i}$ , 
$i=1,...,d$ as belonging to Galois field 
$GF[p]=\{0,1,...,p-1\}$, where $p~is~a~prime~number$ and consider 
the sublattice as a direct product of Galois fields.

$$Z^{d}_{p} = GF[p]\otimes...\otimes GF[p] .$$
As we already mentioned, this reduction of a dynamical system (1)
to a dynamical system for which the Galois field plays 
the role of the phase space makes all operations commutative in the 
sense that 

$$\{ A \{ A ~X \}\} = \{ A^{2}~X  \}, $$
where $\{...\}$ means $mod$ operation.
The commutativity of the multiplication and $(mod)$ operation on the 
Galois sublattice means that the periodic trajectory 

$$\{ A \{A.........\{A~X\}...\}\} = X $$
can be represented in the form 

$$ \{ A^{\tau_{p}}~X \} =X. $$
This equation  allows to understand the relation between
eigenvalues of the matrix $A$ and the period of the trajectories.
Indeed let us consider the eigenvalue problem for the matrix $A$ on
a Galois sublattice

$$ A~X = \lambda~X ,\eqno(6)$$
then the period of the given trajectory $\tau_{p}$ can be 
understood as a 
degree of power on which the $\lambda$ reduces to identity ($mod~~p$)

$$\lambda^{\tau_{p}} = 1~~~~~~~~~~~~~~(mod~~p). \eqno(7)$$
The period of the trajectory 
on a Galois sublattice $GF[p]$ is equal therefore to the 
power $\tau_{p}$ in which the eigenvalue of the matrix $A$ 
reduces to identity. 
It is obvious that the same matrix $A$ will have different 
periods on different Galois fields and that this period 
depends on the given prime number $p$,the dimension  of 
matrices $d$ and the initial vector $X_{0}$.

\section{Eigenvalues of the generator and the period of the trajectories}

Thus the actual value of the period $\tau_{p}$ naturally depends 
on the form of eigenvalues $\lambda$ and of the prime number $p$.
Here we can distinguish different cases:

\vspace{.5cm}

i). The eigenvalue $\lambda$ coincides with one of the elements of the 
Galois field $GF[p]$. In that case the period $\tau_{p}$ 
depends on the fact whether eigenvalue coincides with a primitive 
element of the 
Galois field. All elements of the field $GF[p]$ can be 
constructed as powers of primitive element $g$ and $g^{p-1}=1$.
If the eigenvalue coincides with the primitive element 
of the Galois field , 

$$\lambda = g,~~~~~where~g~is~a~primitive~element~of~GF[p], \eqno(8)$$
then the period is maximal and is equal to 
$\tau_{p} = p-1$
 
$$\lambda^{p-1} =1,~~~~~~~~~(mod~~p). \eqno(9)$$
Therefore to get the maximal  period in the case i)
one should have an eigenvalue equal to the 
primitive element $g$.
If $\lambda$ does not coincide with the primitive element $g$, 
then the period is simply smaller and is equal to $(p-1)/m$
where m is a divisor of $p-1$. 
\vspace{.5cm}

ii). The eigenvalue does not coincide with any 
element of the Galois field $GF[p]$. This may happen 
because Galois field is arithmetically 
complete, but it is not algebraically complete, therefore one 
can have the situation when
the solution of the characteristic polynomial of the 
K-matrix is not in the field $GF[p]$. 
In that case one should ask, whether it is  an 
element of the quadratic extension $GF[\sqrt{p}]$ or of higher 
extensions. The 
quadratic extension of the Galois field consists of the numbers 
of the form $a+b\sqrt g$ where $a,b$ are the elements of field $GF[p]$, 
$g$ is the primitive element of $GF[p]$ and 
$\sqrt g$ is a square-free integer. The primitive element of the 
$GF[\sqrt{p}]$ has the period equal to $p^{2}-1$ \cite{apostol}. 

Thus if the eigenvalue is an element of the quadratic extension and 
coincides with its primitive element $h$

$$\lambda=h,~~~~~where~~~h =
h_{1}+h_{2}\sqrt{ g}~~~~~~~is~a~primitive~element~of~GF[\sqrt{p}], 
\eqno(10)$$  
then the period is equal to $\tau_{p} = p^{2}-1$

$$\lambda^{p^{2}-1} =1,~~~~~~~~~(mod~~p). \eqno(11)$$

\vspace{.5cm}

iii). In general the characteristic polynomial of the K-matrix is 
of order $d$ and the eigenvalue may belong to high extensions 
$GF[\sqrt[d]{p}]$ of the Galois field.  The elements of 
$GF[\sqrt[d]{p}]$ have
the form $a+bh+...+eh^{d-1}$~~where $a,b,...,e$ are the elements of 
$GF[p]$ and  $h$ is a primitive element of $GF[\sqrt[d]{p}]$ 
\cite{lidl,apostol,bastida}.
If the eigenvalue $\lambda$ coincides with this primitive element 

$$\lambda=h~~~~where~h~is~a~primitive~element~of~GF[\sqrt[d]{p}],
\eqno(12)$$  
then the period is equal to $\tau_{p}= p^{d}-1$ \cite{lidl,apostol,bastida}

$$\lambda^{p^{d}-1}=1,~~~~~~~~(mod~~p). \eqno(13)$$

\vspace{.5cm}

{\it This analysis demonstrates an important fact that in order to have a 
large period on a 
sublattice $GF[p]$ one should have K-matrices with 
eigenvalues in high extensions of the field.}

\section{Generators with largest period}

In the previous sections we described the trajectories 
of the K-system on the rational sublattice $Z^{d}_{p}$ and 
particularly 
on a Galois field, that is when p is a prime number.
We have seen that the period of the trajectories
depends on the "order" of the corresponding eigenvalue and 
the period is as large as the order of the extension of the field 
to which belongs the eigenvalue. The question is: 
can we construct a K-matrices with  eigenvalues 
in high extensions of the Galois field and how many of them 
can  {\it simultaneously} belong to a maximal extension $GF[\sqrt[d]{p}]$ ?

We should remark that the d-dimensional 
matrices  $A$ with $all$ eigenvalues in $GF[\sqrt[d]{p}]$ are 
well known in number theory and correspond to so called {\it primitive 
matrices} of the field $GF[\sqrt[d]{p}]$ and  
{\it the determinant of primitive matrices is not equal to one} 
\cite{lidl,apostol,bastida}.
Therefore the  K-matrices which have the determinant equal to one 
can not coincide with 
the primitive matrices, but as we will see one can
construct  K-matrices with $d-1$ eigenvalues in 
$GF[\sqrt[d]{p}]$ and only one in $GF[p]$. This means that
most of the trajectories will have the maximal period 
$\tau_{p}=p^{d}-1$ and only few of them (corresponding to 
that exceptional eigenvalue) 
will have smaller period and we should exclude them from initial 
data.

First let us construct the K-matrices which have the 
eigenvalues in quadratic extension  $GF[\sqrt{p}]$.
If $h$ is the primitive element of the $GF[\sqrt{p}]$, that is 

$$h=h_{1}+h_{2}\sqrt{g},~~~~~~~h\cdot 
h^{\star}=g,~~~~ h+ h^{\star}=2h_{1},\eqno(14)$$
then the matrix which has the eigenvalues in $GF[\sqrt{p}]$ can be 
constructed in the form of (4) 

$$A_{3}= \left( \begin{array}{c}
   0,~~~~~~~~~~~~1,~~~~~~~~~~~~~~~0 \\    
   0,~~~~~~~~~~~~0,~~~~~~~~~~~~~~~1\\ 
   -1,~~~~~~2h_{1}g^{-}-g,~~~~2h_{1}-g^{-}     
\end{array} \right),~~~~(mod~~p) \eqno(15)$$
because the characteristic equation is 

$$(\lambda +g^{-})(\lambda -h)(\lambda-h^{\star})=$$
$$\lambda^{3}-(2h_{1}-g^{-})\lambda^{2}-
(2h_{1}g^{-}-g)\lambda +1=0~~~~(mod~~p) \eqno(16)$$
and has two roots  in $GF[\sqrt{p}]$ and one root in $GF[p]$.
The period  of the most trajectories is equal to 

$$\tau_{p}=p^{2}-1. \eqno(17)$$
and is quadratic in $p$. At the same time 
the trajectories with the initial vector corresponding  
to eigenvalue $\lambda=-g^-$ are  smaller and one should exclude
them from initial data. It is also easy to see that if we 
want to construct two-dimensional K-matrices with 
eigenvalues only 
in $GF[\sqrt{p}]$ we face the problem with determinant
$Det~A =h\cdot h^{\star}=g \neq 1$. This observation explains 
why two-dimensional K-systems, like Arnold cat, can not have 
periodic trajectories of the length $p^2 - 1$ on any Galois
sublattice.

To construct a K-matrix generator with eigenvalues 
in high field $GF[\sqrt[d]{p}]$ we will use primitive 
polynomial of degree $d$ over $GF[\sqrt[d]{p}]$.
The primitive polynomial has the form \cite{apostol,lidl,bastida}

$$\lambda^{d}+\beta_{1} \lambda^{d-1}+\beta_{2}\lambda^{d-2}+...
+\beta_{d} =0\eqno(18)$$
with coefficients $\beta_{1},\beta_{2},...,\beta_{d}$ over $GF[p]$.
The roots of this characteristic polynomial coincide with different
powers of a primitive element $h$ (12) of $GF[\sqrt[d]{p}]$

$$\lambda_{1}=h,~~~~\lambda_{2}=h^{p},~~~...~~~,\lambda_{d}
=h^{p^{d -1}} $$
If $p^{d} -1$ is not divisible by $p,~p^2,...,p^{d -1}$, then all 
of them are primitive elements of $GF[\sqrt[d]{p}]$.
This is the reason why this polynomial is called "primitive".
There are two equivalent representations of $h$: i) in the form of 
root of the  polynomial (18)
and ii) in the form of corresponding matrix \cite{apostol,lidl,bastida}

$$A_{d} =\left( \begin{array}{c}
         0,~~1~,~~0~,...................................,~~0 \\
         0,~~0~,~~1~,...................................,~~0   \\
         .............   \\
         ............. \\
         0,~~~~~~...~~~~~~~~~~0~~~~~~~~~,~~~~~~~~~~~~~~~1~~ \\
         -\beta_{d},~~~~~~.....~~~~~~~,-\beta_{2},~~~~-\beta_{1}
\end{array} \right)~~~~~~(mod~~p) . \eqno(19)$$
As we already explained the problem is that the primitive 
polynomial (18) and the corresponding primitive matrix (19) do not have 
determinant equal to one, because  $\beta_{d} \neq 1$. But this  
property is incompatible with K-condition (2). The exceptional 
case is only $GF[2]$ .

Nevertheless one can solve this 
problem as follows: the last term $\beta_{d}$ which is 
equal to the determinant of the primitive matrix coincides with
the  primitive element 
$g$ of $GF[p]$~~~$\beta_{d}=g$, therefore if we multiply 
the primitive polynomial (18) by $\lambda+g^{-}$ we will get the 
polynomial 

$$(\lambda +g^{-})(\lambda^{d}+\beta_{1} \lambda^{d-1}+
\beta_{2}\lambda^{d-2}+...+\beta_{d})=$$
$$\lambda^{d+1}+(\beta_{1}+g^{-})\lambda^{d}+
(\beta_{2}+\beta_{1}g^{-})\lambda^{d-1}+...
+1=0.\eqno(20)$$
to which corresponds a matrix with unit determinant
of the form (4) 

$$A_{d+1} =\left( \begin{array}{c}
         0,~~1~,~~0~,...................................,~~0 \\
         0,~~0~,~~1~,...................................,~~0   \\
         .............   \\
         ............. \\
         0,~~~~~~...~~~~~~~~~~0~~~~~~~~~,~~~~~~~~~1~~~~~ \\
         -1,.....,-(\beta_{2}+\beta_{1}g^{-}),~~~~-(\beta_{1}+g^{-})
\end{array} \right)~~~~~~(mod~~p) . \eqno(21)$$
of dimension $d+1$. The trajectories generated by this matrix
will have the period

$$\tau_{p}=p^{d}-1 \eqno(22)$$
and we should exclude "dangerous" trajectories corresponding to 
eigenvalue $\lambda=-g^{-}$. They have the form
$X^{(0)}=(x_{1},~ x_{1}(-g^{-}),...,x_{1}(-g^{-})^{d})$ and very short 
period $(p-1)/2$. 

Fascinating result of Niedereiter \cite{niede}
allows to estimate the uniformity of maximally long trajectories 
in terms of Kolmogorov discrepancy $D_{N}$

$$\frac{D_{\tau_{p}}}{\tau_{p}} = \frac{1}{\tau_{p}+1}. $$
The result
is very important because the convergence of the Monte-Carlo simulations
essentially depends on $D_{N}$ \cite{savvidy1}.

The example of the primitive polynomial on $GF[\sqrt[d]{7}]$
with $d=10$ is \cite{lidl}~~~ $\lambda^{10}+\lambda^{9}+
\lambda^{8}+3=0 $~~~~and (20) has the form $\lambda^{11}-
\lambda^{10}-\lambda^{9}-
2\lambda^{8}-4\lambda+1=0$~~~therefore the matrix is 

$$A_{11} =\left( \begin{array}{c}
         0,~~1~,~~0~,.....,~~0 \\
         0,~~0~,~~1~,.....,~~0   \\
         .............   \\
         ............. \\
         0,0,0,...,0,0,0,1 \\
         -1,4,0,...,0,2,1,1
\end{array} \right)~~~~~~(mod~~7) \eqno(23)$$
and the trajectories have the period $7^{10}-1$ except of two trajectories 
with the initial vectors of the form $X^{(0)} = 
(1, 2, 4, 1, 2, 4, 1, 2, 4, 1, 2)$ and 
$X^{(0)} = ( 3, 6, 5, 3, 6, 5, 3, 6, 5, 3, 6 )$. The same matrix will
have different properties on Galois field $GF[p']$ where $p' \neq p$.
The determination of the set of primes for which a given matrix has 
the maximal period is an  unsolved problem \cite{keating}.

Tables of primitive polynomials 
with large values of $d$ are available \cite{lidl}.
In particular \cite{watson}
contains tables for $d<101$, in \cite{stahnke} for $d<169$ and 
in \cite{brillhart} for $d<1001$ with the corresponding periods of order
$2^{1000}$.

\subsection{Conclusion}

In this article we advocate two approaches to generate pseoudorandom 
numbers of high quality: i) the first one is based on K-system generators 
with their exponentially unstable trajectories uniformly filling the phase 
space and ii) on primitive matrices acting on a given Galois sublattice 
with their maximally long trajectories. We demonstrate that one can 
combine these properties in a unique K-matrix which has 
primitive matrix as a submatrix. This construction guarantees that the 
trajectories belong to a K-system and at the same time have maximally large
periods. 

\vspace{1.0 cm}

{\it Acknowledgments} 

We are thankful to N.Akopov for his interest and discussions 
in the early stages of this work.

\vfill
\end{document}